\documentstyle[12pt]{article}

\newcommand{\br}{{\bf r}}
\newcommand{\anu}{{|\nu|}}
\newcommand{\aom}{{|\Omega|}}

\begin{document}
\font\ninerm = cmr9

\def\footnoterule{\kern-3pt \hrule width \hsize \kern2.5pt}

\pagestyle{empty}
\begin{center}
{\large\bf Schr\"odinger Self-adjoint Extension and Quantum Field Theory}%
\footnote{\ninerm This work is supported
in part by funds provided by the U.S. Department of Energy (D.O.E.)
under cooperative agreement \#DE-FC02-94ER40818, as well as in part
by the National Science Foundation under contracts \#INT-910559 and
\#INT-910653, and by Istituto Nazionale di
Fisica Nucleare (INFN, Frascati, Italy). }

\vskip 1cm
Giovanni AMELINO-CAMELIA and Dongsu BAK
\vskip 0.5cm
{\it Center for Theoretical Physics\\
Laboratory for  Nuclear Science and Department of Physics\\
Massachusetts Institute of Technology\\
Cambridge, Massachusetts 02139, U.S.A.}

\end{center}

\vspace{1.2cm}
\begin{center}
{\bf ABSTRACT}
\end{center}

{\leftskip=0.6in \rightskip=0.6in

We argue that the results obtained using
the quantum mechanical method of self-adjoint extension
of the Schr\"odinger Hamiltonian
can also be derived using
Feynman perturbation theory in the investigation of the corresponding
non-relativistic field theories.
We show that this is indeed
what happens in the study of an anyon system, and,
in doing so, we establish a field theoretical
description for ``colliding anyons", {\it i.e.} anyons whose
quantum mechanical wave functions satisfy the non-conventional
boundary conditions obtained with the
method of self-adjoint extension.
We also show that analogous results hold for a system of
non-abelian Chern-Simons particles.

}

\vskip 2cm
\centerline{Submitted to: Physics Letters B}

\vfill

\hbox to \hsize{MIT-CTP-2329 \hfil June 1994}

\newpage
\baselineskip 12pt plus .5pt minus .5pt
\pagenumbering{arabic}
\pagestyle{plain}

\section{Introduction}
The method of self-adjoint extension of the Hamiltonian has been advocated
in several occasions in the literature (see for
example Refs.[1-7]).
In some instances, like in the case of the Dirac Hamiltonian for the
spinning cone in Ref.\cite{desocom}, a
self-adjoint extension of the Hamiltonian (in which
the wave functions are allowed to diverge at a finite number of points,
provided they remain square integrable)
is necessary because the
requirement that the wave functions be regular
does not lead to a self-adjoint Hamiltonian.
For other physical systems, including the ones considered in this Letter,
the demand of regularity
does lead to a self-adjoint Hamiltonian, but one notices
that a one-parameter family of self-adjoint Hamiltonians
(which includes,
in correspondence with a
specific choice of the self-adjoint extension parameter,
the conventional Hamiltonian with regular wave functions)
can be found by relaxing this requirement
(see for example Refs.[1,4-6]).

In the study of
quantum mechanical Schr\"odinger problems\cite{jackbeg,manu,bour}
the results obtained using
the method of self-adjoint extension
raise an interesting issue which has not yet
been addressed in the literature.
It has not been established whether or not
there is a procedure that allows to rederive these results
in the framework of the
field theoretical description (which is supposed to be completely
equivalent to the quantum mechanical description)
of the same non-relativistic
physical system.

In this Letter, we study a system of anyons\cite{wilc} and a system
of non-abelian Chern-Simons (NACS) particles\cite{lee,BJP},
and we prove that the scattering amplitude obtained by a
field theoretical perturbative calculation takes the form of
the corresponding result
obtained using the quantum mechanical method of self-adjoint extension,
provided the renormalized strength of
the contact interaction that is induced
by renormalization\cite{berg,bak}
is chosen to be related in a specific
way (for fixed renormalization scale) to
the self-adjoint extension parameter.

The particular physical systems that we consider
have been recently examined
from different viewpoints in Refs.[11-15],
and are closely related to other extensively
studied problems.
Indeed, anyons, which can be
useful for understanding the
Fractional Quantum Hall Effect\cite{fq},
are particles that acquire fractional statistics through the Aharonov-Bohm
effect\cite{ab},
and our results are therefore related to
the Aharonov-Bohm scattering problem.
Moreover, in Refs.\cite{desocom,desoprd} it was observed that
for an energy eigenstate the equations for a particle in the
gravitational field  of a massless spinning source in two spatial dimensions,
which are also relevant to the study of spinning infinite
cosmic strings in three spatial dimensions,
are equivalent to
those of an infinitely thin flux tube in a background Aharonov-Bohm
gauge field, and are therefore also related to the system here studied.

\section{Abelian Case}
In this section,
we consider anyons, which can be described
as non-relativistic bosons in 2+1 dimensions
interacting through an abelian Chern-Simons gauge field.
The Lagrange density  is
\begin{eqnarray}
 {\cal L}= {\kappa\over 2}\epsilon^{\alpha\beta\gamma}
A_\alpha\partial_\beta A_\gamma
 +i\phi^\dagger D_t \phi -
 {1\over 2}({\bf D}\phi)^\dagger\cdot {\bf D}\phi ~,
\label{a-lag}
\end{eqnarray}
where $\phi$ is a complex
bosonic field, and
$D_t \! \equiv \! \partial_t \! + \! ieA_0$ and
$ {\bf D} \! \equiv \! \nabla \! - \! ie{\bf A}$
are the covariant derivatives.
With the help of the number
operator that is defined as the Noether
charge of the global $U(1)$ symmetry of the Lagrange density,
the quantum theory may be
equivalently formulated as a quantum mechanical
N-body Schr\"odinger problem\cite{JackiwPi}.
In particular, we shall be interested in the
time-independent Schr\"odinger equation describing the 2-body
relative motion:
\begin{eqnarray}
H \psi({\bf r})
\equiv -(\nabla + i\nu \nabla \times \ln r)^2 \psi({\bf r})
= {\bf p}^2 \psi({\bf r}) ~,
\label{a-2body}
\end{eqnarray}
where $\nu={e^2\over 2\pi\kappa}$ is the statistical parameter,
which we can restrict to be in the interval $[-1,1]$ without
loss of generality\cite{wilc}, and
${\bf p}$ is the relative momentum.

If we
demand that the wave functions be regular ({\it i.e.} finite everywhere),
the scattering
problem for the system here considered
is exactly the one solved by Aharonov and Bohm\cite{ab}.
However, in general interesting physical predictions
can also be obtained if the regularity requirement is relaxed, allowing
the wave functions to diverge at a finite number of points,
provided they remain square integrable and the Hamiltonian is
self-adjoint.
Allowing the wave functions $\psi (\br)$ to be non-regular at the
origin ({i.e.} when the particle positions coincide)
leads\cite{manu} to the following one-parameter family
of boundary conditions
at the origin for the s-wave functions:
\begin{equation}
\left[r^\anu\psi(\br) - w R^{2\anu}{d r^\anu \psi (\br)
\over d r^{2\anu}}\right]_{r=0}=0 ~,
\label{bc1}
\end{equation}
which can be equivalently expressed as the following requirement on
the form of $\psi$ for ${\bf r} \sim 0$
\begin{equation}
\psi(\br)\rightarrow a(r^\anu + w R^{2\anu}r^{-\anu})\ \ \
{\rm for} \ r \sim 0 ~.
\label{bc2}
\end{equation}
Here $R$ is a reference scale with dimensions of a length,
$w$ is a dimensionless real parameter,
the self-adjoint extension parameter, which characterizes
the type of boundary condition\footnote{\ninerm Note that,
in Ref.\cite{manu},
the self-adjoint extension is parametrized in terms of a dimensionful
quantity $R_0$, which is related to our $R$ and $w$ by the relation
$(R_0)^{2\anu} = w R^{2\anu}$.}, and $a$ is a constant.

Note that the conformal symmetry possessed\cite{Jackiw}
by the Lagrange density in Eq.(\ref{a-lag}) is in general broken
by the boundary condition (\ref{bc1}) due to the presence of the
dimensionful quantity $w R^{2\anu}$.
Only at the critical points $w \! = \! 0$,
which corresponds to the
conventional Aharonov-Bohm-type scale independent
boundary condition $\psi(0) \! = \! 0$, and $w \! \sim \! \infty$, which
corresponds to the scale independent
boundary condition ${d r^\anu \psi (\br)
\over d r^{2\anu}}\mid_{r=0}= \! 0$, the scale symmetry is preserved.

One can easily see\cite{manu} that for non-s-wave functions
square integrability is only consistent with the $\psi(0)=0$
boundary condition; thus, the method of self-adjoint extension only affects
the s-wave part of the calculations, which are therefore the ones we shall
be concerned with.

The s-wave scattering amplitude for anyons satisfying the
boundary condition (\ref{bc1}) can be evaluated exactly by using
a rather straightforward generalization of the analysis given
in Ref.\cite{ab}, which concerned the special case $w=0$;
we find (also see Ref.\cite{manu})
\begin{eqnarray}
&A_s(p) = - i {\sqrt{{2 \over \pi p}}}
(e^{i\pi\anu} -1){1-{1 \over w}
\left({2\over pR}\right)^{2\anu}{\Gamma(1+\anu)\over
\Gamma(1-\anu)}\over 1+ {1 \over w}
e^{i\pi\anu}\left({2\over pR}\right)^{2\anu}{\Gamma(1+\anu)\over
\Gamma(1-\anu)}} ~~~~~~~~~~ ~~~~~~~~~~~~&
\nonumber\\
&= -{\sqrt{{2\pi \over p}}} \{ \anu {1-w\over 1+w} -{i \pi \over 2}\nu^2
- \nu^2{4w\over (1+w)^2}\left(\ln {pR\over 2}+\gamma -
{i\pi\over 2}\right)   \nonumber\\
&~~~~~~ ~~~~~~~~~~~~ ~~~~\,
- {\pi^2 \over 6}\anu^3 {1-w\over 1+w}
\!-\! \anu^3{4(1-w)w\over (1+w)^3}\left(\ln {pR\over 2}\!+\!
\gamma\! -\!{i\pi\over 2}\right)^2\!\! +\! O(\nu^4)
\} ~,&
\label{a-per1}
\end{eqnarray}
where $p \equiv |{\bf p}|$, and
$\gamma$ denotes the Euler constant.

Note that scale invariance is
in general broken by the
dependence of $A_s$ on $R$.
Consistently with our preceding observation, the scale symmetry
is only preserved at the critical values of $w$, for which
the s-wave amplitude can be written as
\begin{eqnarray}
 A_s = - i {\sqrt{{2 \over \pi p}}}
(e^{\mp i\pi\anu} -1) ~,
\end{eqnarray}
where the upper (lower) sign holds for the $w \! = \! 0$
($w \! \sim \! \infty$) critical point.

As stated in the Introduction, we intend
to show that it is possible to rederive
the quantum mechanical result
for the scattering amplitude obtained using the method of
self-adjoint extension of the Schr\"odinger
Hamiltonian in a field theoretical perturbative calculation.

The field theoretical description of the system that we are considering
has been discussed in Refs.\cite{berg,bak}. It was shown that
renormalizability requires the addition of
a contact term $-\pi g_b (\phi^\dagger\phi)^2$
to the Lagrangian density ${\cal L}$ of Eq.(\ref{a-lag}).
The two-particle scattering amplitude was calculated to one-loop order
in Ref.\cite{bak}; its s-wave part\footnote{Note that in the
field theoretical calculations one can easily show (to all orders)
that the non-s-wave part of the scattering amplitude is cut-off-independent
(and therefore it plays no role in the renormalization procedure),
contact-coupling-independent, and, besides the overall kinematic factor,
scale-independent.} (including the appropriate kinematic factor)
can be written as
\begin{eqnarray}
&A_{s,1-l}(p)=-{\sqrt{{2 \pi \over p}}}\left\{
g_b \! - \! {i \pi \over 2} \nu^2 \!
+ \! (g_b^2 \! - \! \nu^2)
\left(\ln {p\over {\mu'}} \!
- \! {i\pi\over 2} \! - \! {1\over 2\epsilon} \right)
\! - \! g_b^2 \epsilon \left[
\left(\ln {p\over {\mu'}} \!
- \! {i\pi\over 2} \! \right)^2 + {\pi^2 \over 24} \right]
\right\}& \nonumber\\
&=-{\sqrt{{2\pi \over p}}} \left[ g_r - {i \pi \over 2} \nu^2
+ (g_r^2 -\nu^2)\left(\ln {p\over {\mu}}
-{i\pi\over 2}\right) \right] ~,~~~~~~~~~~~~~~~~~~ ~~ &
\label{a-renamp}
\end{eqnarray}
%
where $\epsilon$ is the usual cut-off used in dimensional regularization,
$\mu$ is the renormalization scale,
$\mu'$, which we introduced just
in order to simplify the notation, is defined
by $\ln \mu'= \ln \mu -{\gamma -\ln 4\pi\over 2}$,
and we also introduced a
one-loop renormalized coupling $g_r$ defined in terms of the
bare coupling $g_b$ by the relation
\begin{eqnarray}
&g_r = g_b -
(g_b^2-\nu^2)({1\over 2\epsilon} -
{\gamma -\ln 4\pi\over 2}) ~. &
\label{a-ren}
\end{eqnarray}

Note that only at the critical values $g_r=\pm\anu$ of the renormalized
contact coupling the
scale invariance of the classical thaory is preserved at the
quantum (one-loop) level.
Moreover, it was observed in Refs.\cite{berg,bak}
that at the repulsive
critical value of the renormalized contact
coupling, {\it i.e.} $g_r=\anu$,
the result (\ref{a-renamp}) is consistent
with the Aharonov-Bohm scattering amplitude (which is given by
the $w \rightarrow 0$ limit of Eq.(\ref{a-per1})),
and it is in this sense that this field theory for $g_r=\anu$
describes the conventional anyons, which satisfy the
Aharonov-Bohm-type regular boundary condition.

Our objective is to establish a general connection
(as we just mentioned,
this connection was only understood for the special
case $w=0$, $g_r=\anu$)
between the
$g_r$-dependent field theoretical
results and the $w$-dependent quantum mechanical results of the
method of self-adjoint extension.
We identify this connection by comparison of the
results in Eqs.(\ref{a-per1})
and (\ref{a-renamp}); in fact, we observe that if one uses the relations
\begin{eqnarray}
g_r= \anu {1-w \over 1+w} ~,~~~~\mu = {2\over Re^{\gamma}}
\label{a-rel1}
\end{eqnarray}
the one-loop field theoretical
result (\ref{a-renamp}) reproduces exactly the $O(\nu^2)$
approximation
of the quantum mechanical result (\ref{a-per1}) obtained using the
method of self-adjoint extension.

We observe that, in particular, Eq.(\ref{a-rel1}) implies
that (as it should be expected based on the analysis of scale invariance),
like $w = 0 $ corresponds to $g_r = \anu$,
the other critical value ($w\sim \infty$)
of the self-adjoint extension parameter corresponds to
attractive critical strength\footnote{Note that, in Ref.\cite{JackiwPi}
it was shown that
for attractive critical value of the contact coupling
the classical version of the
field theory here considered admits static solutions, {\it i.e.} solitons,
that satisfy a self-dual equation which is equivalent to the Liouville
equation.
It would be interesting to investigate whether characteristic structures also
arise at the quantum mechanical level  in correspondence of $w \sim \infty$.}
($g_r = -\anu$) of the contact interaction.

Further insight into the correspondence (\ref{a-rel1})
between the quantum mechanical
variables $w$,$R$ and the field theoretical variables $g_r$,$\mu$
can be gained from the following observations.
First, we notice that using
the renormalization-group
equation which states that the physical
scattering amplitude is independent on the
choice of the renormalization scale $\mu$, one can derive,
to the order $\nu^2$, the following beta function for the
coupling $g_r$:
\begin{eqnarray}
\beta(g_r)\equiv {d g_r\over d \ln \mu} = g_r^2 -\nu^2 ~.
\label{a-beta}
\end{eqnarray}
Eq.(\ref{a-beta}), which indicates that $g_r$ and $\mu$ are not physically
independent, can be integrated to give the relation
\begin{eqnarray}
 {\anu +g_r(\mu_1) \over \anu -g_r(\mu_1)}\,\mu_1^{2\anu}
={\anu +g_r(\mu_2) \over \anu -g_r(\mu_2)}\,\mu_2^{2\anu} ~.
\label{a-beta1}
\end{eqnarray}
Similarly in the exact result (\ref{a-per1}), which was obtained in the
quantum mechanical framework, $R$ is only a reference scale, and obviously
physics must be independent of the choice of $R$. Indeed, all physical
quantities (see, for example, Eqs.(\ref{bc1}) and (\ref{a-per1}))
depend on $w$ and $R$ only through the quantity $w R^{2\anu}$,
and the independence of physics on the choice of $R$ is realized
by the fact that if $R$ is changed from a value $R_1$ to a value $R_2$
this must be accompanied by a corresponding change of $w$ as described by the
relation
\begin{eqnarray}
w(R_1) \, R_1^{2\anu} = w(R_2) \, R_2^{2\anu} ~.
\label{a-beta2}
\end{eqnarray}
Clearly, the Eqs.(\ref{a-beta1}) and (\ref{a-beta2}) are perfectly consistent
with the relations (\ref{a-rel1}).
This observation is even more remarkable considering that Eq.(\ref{a-beta2})
is exact, whereas the Eqs.(\ref{a-rel1}) and (\ref{a-beta1}) are just based on
a one-loop analysis.
Evidently, the Eqs.(\ref{a-rel1}) and (\ref{a-beta1})
have more general validity than one would expect based on the fact that
they have been derived at one-loop.
In order to test whether indeed
the Eqs.(\ref{a-rel1}) and (\ref{a-beta1}) receive vanishing
higher loop contributions, we now calculate
the two-loop s-wave scattering amplitude.
The computation is simplified by the fact that
it is easy to show that the only two-loop contributions to the
s-wave scattering
amplitude come from the two diagrams in Fig.1.

\font\ninerm = cmr9
{\ninerm
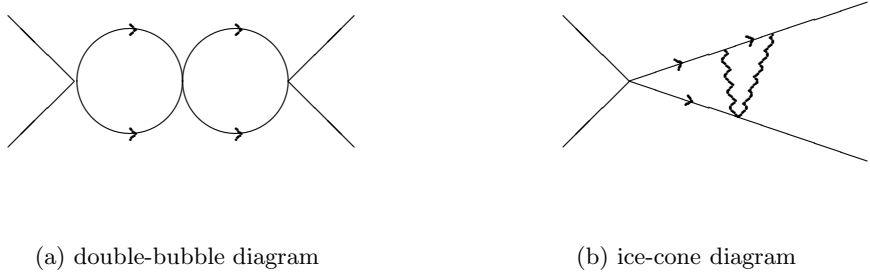
\begin{figure}[h]
\begin{picture}(400, 142)
\put(65,5){\makebox(0,0)[l]{(a) double-bubble diagram}}
\put(270,5){\makebox(0,0)[l]{(b) ice-cone diagram}}
\put(80,72){\line(-1,1){25}}

\def\unused{\put(67.5,86){.}
\put(67.5,85){.}
\put(67.5,84){.}
\put(67.5,83){.}
\put(66.5,83){.}
\put(65.5,83){.}
\put(64.5,83){.}
}
\put(80,72){\line(-1,-1){25}}

\def\unused{\put(67.5,57.5){.}
\put(67.5,58.5){.}
\put(67.5,59.5){.}
\put(67.5,60.5){.}
\put(66.5,60.5){.}
\put(65.5,60.5){.}
\put(64.5,60.5){.}
}
\put(101,72){\circle{60}}
\put(141,72){\circle{60}}
\put(140,93){.}
\put(140.5,92.5){.}
\put(141,92){.}
\put(141.5,91.5){.}
\put(142,91){.}
\put(141.5,90.5){.}
\put(141,90){.}
\put(140.5,89.5){.}
\put(140,89){.}

\put(140,53.5){.}
\put(140.5,53){.}
\put(141,52.5){.}
\put(141.5,52){.}
\put(142,51.5){.}
\put(141.5,51){.}
\put(141,50.5){.}
\put(140.5,50){.}
\put(140,49.5){.}

\put(100,93){.}
\put(100.5,92.5){.}
\put(101,92){.}
\put(101.5,91.5){.}
\put(102,91){.}
\put(101.5,90.5){.}
\put(101,90){.}
\put(100.5,89.5){.}
\put(100,89){.}

\put(100,53.5){.}
\put(100.5,53){.}
\put(101,52.5){.}
\put(101.5,52){.}
\put(102,51.5){.}
\put(101.5,51){.}
\put(101,50.5){.}
\put(100.5,50){.}
\put(100,49.5){.}

\put(161,72){\line(1,-1){25}}

\def\unused{\put(171.5,63.5){.}
\put(171.5,62.5){.}
\put(171.5,61.5){.}
\put(171.5,60.5){.}
\put(170.5,60.5){.}
\put(169.5,60.5){.}
\put(168.5,60.5){.}
}
\put(161,72){\line(1,1){25}}

\def\unused{\put(171.5,80){.}
\put(171.5,81){.}
\put(171.5,82){.}
\put(171.5,83){.}
\put(170.5,83){.}
\put(169.5,83){.}
\put(168.5,83){.}
}
\put(290,72){\line(-1,1){25}}

\def\unused{\put(277.5,86){.}
\put(277.5,85){.}
\put(277.5,84){.}
\put(277.5,83){.}
\put(276.5,83){.}
\put(275.5,83){.}
\put(274.5,83){.}
}
\put(290,72){\line(-1,-1){25}}

\def\unused{\put(277.5,57.5){.}
\put(277.5,58.5){.}
\put(277.5,59.5){.}
\put(277.5,60.5){.}
\put(276.5,60.5){.}
\put(275.5,60.5){.}
\put(274.5,60.5){.}
}

\put(290,72){\line(3,1){90}}
\put(290,72){\line(3,-1){90}}

\put(307.4,76){.}

\put(307.7,76.5){.}
\put(307.9,77){.}

\put(308.1,77.5){.}
\put(308.3,78){.}

\put(308.5,78.5){.}
\put(308,78.7){.}

\put(307.5,78.9){.}
\put(307,79.1){.}

\put(306.5,79.3){.}
\put(306,79.5){.}


\put(311,66.1){.}

\put(311.3,65.8){.}
\put(311.6,65.3){.}

\put(311.9,64.8){.}
\put(312.2,64.3){.}

\put(312.5,63.8){.}
\put(312.0,63.6){.}

\put(311.5,63.4){.}
\put(311,63.2){.}

\put(310.5,62.9){.}
\put(310,62.6){.}


\put(334.4,85){.}

\put(334.7,85.5){.}
\put(334.9,86){.}

\put(335.1,86.5){.}
\put(335.3,87){.}

\put(335.5,87.5){.}
\put(335,87.7){.}

\put(334.5,87.9){.}
\put(334,88.1){.}

\put(333.5,88.3){.}
\put(333,88.5){.}


\put(330,58){.}
\put(329.5,58.5){.}
\put(329,59){.}
\put(328.5,59.5){.}
\put(328,60){.}
\put(327.5,60.5){.}

\put(327,61){.}
\put(327,61.5){.}
\put(327,62){.}
\put(327.5,62.5){.}
\put(328,63){.}
\put(328.5,63.5){.}
\put(329,64){.}
\put(328.5,64.5){.}
\put(328,65){.}
\put(327.5,65.5){.}
\put(327,66){.}
\put(326.5,66.5){.}
\put(326,67){.}
\put(326,67.5){.}
\put(326,68){.}
\put(326.5,68.5){.}
\put(327,69){.}
\put(327.5,69.5){.}
\put(328,70){.}
\put(327.5,70.5){.}
\put(327,71){.}
\put(326.5,71.5){.}
\put(326,72){.}
\put(325.5,72.5){.}
\put(325,73){.}
\put(325,73.5){.}
\put(325,74){.}
\put(325.5,74.5){.}
\put(326,75){.}
\put(326.5,75.5){.}
\put(327,76){.}
\put(326.5,76.5){.}
\put(326,77){.}
\put(325.5,77.5){.}
\put(325,78){.}
\put(324.5,78.5){.}
\put(324,79){.}
\put(324,79.5){.}
\put(324,80){.}
\put(324.5,80.5){.}
\put(325,81){.}
\put(325.5,81.5){.}
\put(326,82){.}
\put(325.5,82.5){.}
\put(325,83){.}

\put(330,58){.}
\put(330.5,58.5){.}
\put(331,59){.}
\put(331.5,59.5){.}
\put(332,60){.}
\put(332.5,60.5){.}
\put(333,61){.}
\put(333,61.5){.}
\put(333,62){.}
\put(332.5,62.5){.}
\put(332,63){.}
\put(332.5,63.5){.}
\put(333,64){.}
\put(333.5,64.5){.}
\put(334,65){.}
\put(334.5,65.5){.}
\put(335,66){.}
\put(335,66.5){.}
\put(335,67){.}
\put(334.5,67.5){.}
\put(334,68){.}
\put(334.5,68.5){.}
\put(335,69){.}
\put(335.5,69.5){.}
\put(336,70){.}
\put(336.5,70.5){.}
\put(337,71){.}
\put(337,71.5){.}
\put(337,72){.}
\put(336.5,72.5){.}
\put(336,73){.}
\put(336.5,73.5){.}

\put(337,74){.}
\put(337.5,74.5){.}
\put(338,75){.}
\put(338.5,75.5){.}
\put(339,76){.}
\put(339,76.5){.}
\put(339,77){.}
\put(338.5,77.5){.}
\put(338,78){.}
\put(338.5,78.5){.}
\put(339,79){.}
\put(339.5,79.5){.}
\put(340,80){.}
\put(340.5,80.5){.}
\put(341,81){.}
\put(341,81.5){.}
\put(341,82){.}
\put(340.5,82.5){.}
\put(340,83){.}
\put(340.5,83.5){.}
\put(341,84){.}
\put(341.5,84.5){.}
\put(342,85){.}
\put(342.5,85.5){.}
\put(343,86){.}
\put(343,86.5){.}
\put(343,87){.}
\put(342.5,87.5){.}
\put(342,88){.}
\put(342.5,88.5){.}
\put(343,89){.}

\end{picture}
\caption{ The two-loop diagrams that contribute to the s-wave scattering
amplitude.\ }
\end{figure}
}

\noindent
We find that the contribution $A_{db}$ of the double-bubble diagram (Fig.1a)
and the contribution $A_{ic}$ of the ice-cone diagram (Fig.1b)
are given by
\begin{eqnarray}
&A_{db}
= - {\sqrt{{2\pi \over p}}} g_b^3
\left[{1\over 4\epsilon^2}-{1\over\epsilon}
\left(\ln {p\over {\mu'}} -{i\pi\over 2} \right) +
2\left(\ln {p\over {\mu'}} -{i\pi\over 2} \right)^2 + {\pi^2\over 24}\right]
{}~, &
\label{a-db}\\
&A_{ic}
= {\sqrt{{2\pi \over p}}} g_b \nu^2
\left[{1\over 4\epsilon^2}-{1\over\epsilon}
\left(\ln {p\over {\mu'}} -{i\pi\over 2} \right) +
2\left(\ln {p\over {\mu'}} -{i\pi\over 2} \right)^2
+ {5 \pi^2\over 24}\right] ~. &
\label{a-ic}
\end{eqnarray}
Adding $A_{db}$ and $A_{ic}$ to the one-loop
s-wave scattering amplitude, and introducing a two-loop renormalized
contact coupling $g_r$ related to the bare coupling $g_b$ by
\begin{eqnarray}
&g_r = g_b - (g_b^2-\nu^2)({1\over 2\epsilon} - {\gamma -\ln 4\pi\over 2})
+g_b(g_b^2-\nu^2)({1\over 2\epsilon} - {\gamma -\ln 4\pi\over 2})^2 ~,&
\label{a-reno}
\end{eqnarray}
we obtain the following final result for the renormalized s-wave
scattering amplitude to two-loop order:
\begin{eqnarray}
&A_{s,2-l} \! = \! -{\sqrt{{2\pi \over p}}} \{
g_r \!-\! {i \pi \over 2} \nu^2 \!-\! {\pi^2 \over 6} g_r \nu^2
\!+\! (g_r^2 \!-\! \nu^2)
\! \left(\ln {p\over {\mu}}
\!-\! {i\pi\over 2} \right)
\!+\! g_r(g_r^2 \!-\! \nu^2)
\! \left(\ln {p\over {\mu}} \!-\! {i\pi\over 2} \right)^2 \} .\,~~&
\label{a-peramp2}
\end{eqnarray}
It is easy to verify that
this result reproduces the $O(\nu^3)$ approximation
of the exact result (\ref{a-per1})
if one uses again the  relations (\ref{a-rel1}); therefore,
we find  that, as anticipated by our hypothesis that
these relations are exact, there is no two-loop correction
to the relations (\ref{a-rel1}).
Analogously, from Eq.(\ref{a-peramp2}) one can verify that there is no
two-loop correction to the $\beta$-function given in Eq.(\ref{a-beta}).
(Using a different formalism, in Ref.\cite{loza} it was shown
that also the three-loop correction to Eq.(\ref{a-beta}) vanishes.)


\section{Non-abelian Generalization}

In this section, we discuss the generalization of
the analysis presented in the preceding section to the case of
a non-abelian gauge-symmetry group.
The system considered is therefore exactly the one of Ref.\cite{bak},
and we adopt the notation and conventions introduced in that paper,
with the only exception that, for simplicity, we study
particles of unit mass.


The non-abelian generalization of the Schr\"odinger problem (\ref{a-2body})
is given by\cite{lee,BJP}
\begin{eqnarray}
H \Psi({\bf r}) \equiv -(\nabla + i\Omega \nabla \times \ln r)^2
\Psi({\bf r}) = {\bf p}^2 ~,
\label{na-2body}
\end{eqnarray}
where $\Omega = -{g^2\over 2\pi\kappa} T^a \otimes T_a$,
and $\Psi =\sum_{nm}\psi_{nm}(\br) |nm\rangle $ is a two-NACS-particle
state (the $\psi_{nm}$ are the components of $\Psi$ in the
$|nm\rangle$ basis, see Ref.\cite{bak}).

By choosing a basis which diagonalizes
$\Omega$ this non-abelian problem can be essentially reduced to the
abelian one. Exploiting this simplification, it is easy to show that in
the non-abelian case the method of self-adjoint extension of the
Schr\"odinger Hamiltonian leads
to the following requirement on the form of the s-wave functions
for $r \sim 0$:
\begin{equation}
\Psi(\br) \rightarrow
(r^\aom +r^{-\aom}R^{\aom} W R^{\aom})\sum_{nm} a_{nm} |nm\rangle
\ \ \ {\rm for} \ r \sim 0 ~,
\label{na-bc}
\end{equation}
where the $a_{nm}$ are constants, $W$ is a Hermitian
matrix\footnote{We observe
that for a general Hermitian matrix the form (\ref{na-bc}) of the s-wave
functions is not gauge covariant. Gauge covariance can be achieved by
demanding that $W$ satisfies the relation
$[W, T_a \! \otimes 1 \! + \! 1 \! \otimes \! T_a]=0$,
which also implies that $W$ commutes with $\aom$.
For completeness, we present formulas valid for a
general Hermitian matrix $W$, {\it i.e.} we keep track of
the ordering of the matrices $W$ and $\aom$.}
whose components are dimensionless parameters that characterize the
self-adjoint extension, and
$|\Omega|$ is the matrix which in the basis that diagonalizes $\Omega$
has elements given by the absolute value of the elements of $\Omega$.

In a basis that diagonalizes $\Omega$, it is also easy to obtain the
non-abelian generalization of Eq.(\ref{a-per1}).
We find that the boundary condition (\ref{na-bc}) leads to an s-wave
amplitude for the scattering $n_1,m_1 \! \rightarrow \! n_2,m_2$
which is given by $\langle n_2, m_2|{\cal A}_s|n_1, m_1\rangle +
\langle m_2, n_2|{\cal A}_s|n_1, m_1\rangle$ with
\begin{eqnarray}
&{\cal A}_s \!\! = \!\! - {i \over {\sqrt{2 \pi p}}}
(e^{i\pi\aom} -1)
\{1-e^{-{i\pi\aom\over2}}\left({2\over pR}\right)^{\aom}\Gamma(1+\aom)
{1 \over W}
{1\over \Gamma(1-\aom)}
\left({2\over pR}\right)^{\aom}e^{{i\pi\aom\over2}}\}&
\nonumber\\
& ~~ ~~~~~~~~~~~~~~~~~~~~~~~~~~~~ \cdot \{1+e^{{i\pi\aom\over2}}
\left({2\over pR}\right)^{\aom}\Gamma(1+\aom)
{1 \over W}
{1\over \Gamma(1-\aom)}
\left({2\over pR}\right)^{\aom}e^{{i\pi\aom\over2}}\}^{-1} ~. &
\label{na-per1}
\end{eqnarray}

%
%
%

As done in the preceding section for the abelian case,
we now proceed to establish a prescription that allows to rederive this
result, which was obtained using the method of self-adjoint extension,
in the framework of the field theoretical technique of Feynman
perturbation theory.
The renormalized scattering amplitude was calculated in field theory
to one-loop order
in Ref.\cite{bak}; its s-wave part is
\begin{eqnarray}
&{\cal A}_{s,1-l} = -{\sqrt{{\pi \over 2 p}}}
\left[G_r -{i \pi \over 2} \Omega^2
+ (G_r^2 -\Omega^2)\left(\ln {p\over {\mu}} -{i\pi\over 2}\right)
\right] ~, &
\label{na-peramp1}
\end{eqnarray}
where $G_r$ is the renormalized contact coupling matrix that appears in
the contact term required for renormalizability of the non-abelian model
(specifically, $G_r$ is defined in terms of the matrix $C_r$ introduced
in Ref.\cite{bak} by the relation $G_r \!=\! C_r/4 \pi$).
We observe that Eq.(\ref{na-peramp1})  reproduces the $O(|\Omega|^2)$
approximation of Eq.(\ref{na-per1}) if one uses the following relations
\begin{eqnarray}
G_r = \aom (1-W) (1+W)^{-1} ~,~~~~
\mu = {2\over R e^{\gamma}} ~.
\label{na-rel2}
\end{eqnarray}

Also for this non-abelian case, in order to present some evidence that
the relations (\ref{na-rel2}) are stable with respect to higher order
contributions, we calculated
the renormalized two-loop
s-wave scattering amplitude; our result is
\begin{eqnarray}
&{\cal A}_{s,2-l} =-{\sqrt{{\pi \over 2 p}}}
\{G_r - {i \pi \over 2} \Omega^2
- {\pi^2 \over 12} (G_r \Omega^2 + \Omega^2 G_r)
+ (G_r^2 -\Omega^2)\left(\ln {p\over \mu} -{i\pi\over 2}\right) &
\nonumber\\
&~~~~ ~~~~~~~ + {1 \over 2}( G_r(G_r^2 -\Omega^2)+(G_r^2 -\Omega^2)G_r)
\left(\ln {p\over \mu} -{i\pi\over 2}\right)^2 \} ~, &
\label{na-per2}
\end{eqnarray}
which does indeed reproduce the $O(|\Omega|^3)$
approximation of Eq.(\ref{na-per1}) when the relations (\ref{na-rel2})
are used.

\section{Conclusion}
\renewcommand{\theequation}{\Roman{section}.\arabic{equation}}
In our study of anyons and of NACS particles, we have identified relations
between the quantities that appear in the (quantum mechanical)
method of self-adjoint extension and the
quantities that appear in the (field theoretical) Feynman perturbation theory,
which allow to put in one to one correspondence the results of the
two methods.
In establishing this result, we have extended to
two-loop (see Eqs.(\ref{a-peramp2}) and (\ref{na-per2})) some of the
results of Refs.\cite{berg,bak}, and we have generalized to the case of
NACS particles (see Eqs.(\ref{na-bc})) the results of the
method of self-adjoint extension presented
for anyons in Refs.\cite{manu,bour}.

Based on our analysis, we argue that
in general the results obtained using
the method of self-adjoint extension of the Schr\"odinger Hamiltonian
should be equivalently derivable
by a (suitably renormalized) perturbative calculation
in the framework of the corresponding field theoretical problem.

A problem of theoretical physics that is related to the ones here studied,
is the connection between boundary conditions for the wave functions
and contact interactions, which was recently investigated in
quantum mechanics\cite{jackbeg,amelplb,ouvr}.
The analysis presented in the preceding sections  shows that also in
field theory the introduction of contact interactions can be used
to implement
in the perturbative calculations
a choice of boundary conditions for the wave functions.

Our investigation is also relevant to the issue
of which boundary conditions at the points of overlap of particle positions
are most natural in the case of anyons\cite{bour,amelprl} or NACS particles.
The results of Sec.II (Sec.III)
establish a field theoretical description of ``colliding anyons"\cite{bour}
(``colliding NACS particles"), {\it i.e.} we identified the strength
of the contact coupling $g_r$ which (at fixed renormalization scale)
is to be used in the field theory calculations to describe anyons
(NACS particles) whose
quantum mechanical wave functions satisfy the non-conventional
boundary conditions obtainable with the
method of self-adjoint extension with parameter $w$ (parameter matrix $W$).
In this field theoretical formalism there appears to be no
reason
for restricting oneself
to the case of the conventional ``non-colliding anyons"\cite{bour},
the ones whose wave functions vanish
at the points of overlap of particle positions\footnote{The
only special property of the
non-colliding anyons, which correspond to $w = 0$,
is the preservation of the scale invariance; however, this property
is shared by the case of colliding anyons with $w \sim \infty$,
and, anyway, it is not clear to us whether
there is any physical motivation to
exclude values of $w$ that do not preserve scale invariance.}.

Lastly, we want to point out that the calculations
presented in this Letter give one of the rare opportunities of
comparing exact results to renormalization requiring perturbative
field theoretical results, and we hope they can be used to gain
some insight in the
physics behind the regularization and renormalization procedure.
For example,
in our analysis it appears that the necessity of a cut-off
is not a relict of some unknown
ultraviolet physics, but rather an artifact of the
perturbative methods used.
This is in contrast with the conventional wisdom
on renormalization; however, our results are derived in the context of
non-relativistic field theory,
and it is unclear to us whether similar conclusions
could be reached in the case of relativistic field theories,
which is the framework where renormalization is customarily used.

\bigskip
\bigskip
\ \indent
The authors would like to thank R. Jackiw for
useful comments and for critical reading of the manuscript, and
to acknowledge discussions on the subject of this paper with O. Bergman
and C. Manuel.
\newpage
\baselineskip 12pt plus .5pt minus .5pt

\end{document}